\documentclass[aps,prl,amsmath,amssymb,reprint,twocolumn,superscriptaddress,showpacs]{revtex4}
\usepackage{epsfig,amssymb}
\usepackage{graphicx}
\usepackage{dcolumn}
\usepackage{bm}
\usepackage{dcolumn}
\usepackage{multirow}
\begin{document}
\title{Effect of spin fluctuations on quasiparticle excitations: first-principles
theory and application to sodium and lithium}

\author{Johannes~Lischner}
\email{jlischner@civet.berkeley.edu}
\affiliation{Department of Physics, University of California,
  Berkeley, California 94720, USA, and Materials Sciences Division,
  Lawrence Berkeley National Laboratory, Berkeley 94720, USA.}
\author{Timur~Bazhirov}
\affiliation{Department of Physics, University of California,
  Berkeley, California 94720, USA, and Materials Sciences Division,
  Lawrence Berkeley National Laboratory, Berkeley 94720, USA.}
\author{Allan H. MacDonald}
\affiliation{Department of Physics, The University of Texas at Austin,
  Austin, Texas, 78712, USA.}
\author{Marvin L. Cohen}
\affiliation{Department of Physics, University of California,
  Berkeley, California 94720, USA, and Materials Sciences Division,
  Lawrence Berkeley National Laboratory, Berkeley 94720, USA.}
\author{Steven G. Louie}
\affiliation{Department of Physics, University of California,
  Berkeley, California 94720, USA, and Materials Sciences Division,
  Lawrence Berkeley National Laboratory, Berkeley 94720, USA.}

\begin{abstract}
  We present first-principles calculations for quasiparticle
  excitations in sodium and lithium including the effects of charge
  and spin fluctuations. We employ the Overhauser-Kukkonen form for
  the electron self energy arising from spin fluctuations and
  demonstrate that the coupling of electrons to spin fluctuations
  gives an important contribution to the quasiparticle lifetime, but
  does not significantly reduce the occupied bandwidth. Including
  correlation effects beyond the random-phase approximation in the
  screening from charge fluctuations yields good agreement with
  experiment.
\end{abstract}

\maketitle


\emph{Introduction}.---The coupling of electrons to spin fluctuations
causes many fascinating phenomena: for example, it has been proposed
that spin fluctuations can ``glue'' electrons together to form Cooper
pairs giving rise to unconventional high-temperature superconductivity
\cite{Keimer,Keimer2,Pines,Scalapino}. In particular, spin
fluctuations were invoked to explain superconductivity in the cuprates
\cite{Pines,Keimer,Keimer2} and recently also in the iron pnictide and
chalcogenide materials \cite{Mazin,Aoki,Graser}. In addition, it is
well known that the coupling of spin fluctuations to electrons can
affect the electronic effective mass and consequently transport
properties and the specific heat.

Theoretically, the effect of spin fluctuations on quasiparticle
excitations is usually calculated using model Hamiltonians. Early
studies \cite{Doniach,BerkSchrieffer,Schrieffer} constructed empirical
theories including spin fluctuations based on the homogeneous electron
gas and simple tight-binding models. More recently, many empirical
theories involving spin fluctuations were constructed to investigate
superconductivity in the cuprates and pnictides. In these theories the
spin susceptibility is either parametrized using experimental neutron
scattering and nuclear magnetic resonance data
\cite{Pines,Eschrig,Keimer} or estimated by combining
density-functional theory (DFT) with interaction parameters (such as
the Hubbard $U$) adjusted to reproduce experimental
findings\cite{Graser,Aoki}.

While the aforementioned theories have been very instructive, their
applications have been limited by the availability of concrete
experimental data needed to determine their input parameters,
supporting the need for a fully first-principles theory without
empirical parameters.  There have been several attempts to compute the
spin fluctuation-electron coupling from first principles. Notably,
Winter and coworkers \cite{Winter1,Winter2} calculated the spin
susceptibility and the spin fluctuation-electron self energy from DFT
and evaluated the correction to the specific heat for palladium and
vanadium. Later studies \cite{Aryasetiawan,Echenique1,Echenique2}
employed a first-principles T-matrix approach to calculate satellites
in the photoemission spectrum of nickel and quasiparticle lifetimes in
metals. However, such theory requires the solution of a
computationally expensive four-point equation and also a correction to
account for the double counting of certain Feynman diagrams.

The alkali metals are arguably among the simplest systems in condensed
matter theory. Careful comparisons between theory and experiment in
these systems has guided progress in understanding electron
correlation physics in the itinerant-electron limit. In particular,
the Fermi surfaces of sodium and lithium are highly spherical
indicating that a description based on the homogeneous electron gas
might be valid. Therefore it was surprising when angle-resolved
photoemission experiments \cite{Plummer1,Plummer2} reported a
substantially smaller occupied bandwidth than was found in Hartree
calculations on the homogeneous electron gas and also in DFT
calculations. In addition, self-energy corrections employing the
standard GW approximation to the electron self energy
\cite{LouieHybertsen,HedinBook}, where the self energy is expressed as
the product of the interacting Green's function $G$ and the screened
Coulomb interaction $W$, could not account for the full
reduction of the occupied bandwidth indicating that electron
correlation effects not included in these calculations play an
essential role in these materials.

Northrup, Hybertsen, and Louie \cite{Northrup1,Northrup2} included
vertex corrections in the dielectric matrix approximately by computing
the charge susceptibility from DFT instead of employing the
random-phase approximation (RPA) and found the resulting GW values in
good agreement with experiment for the occupied bandwidths of lithium
and sodium. At the same time, Zhu and Overhauser \cite{OverhauserZhu}
found that spin fluctuations within a paramagnon pole model could also
explain the reduction of the bandwidth. This difference in the
mechanism responsible for the band width reduction in the alkali
metals has not yet been resolved. In particular, no first-principles
calculation of the self-energy correction arising from spin
fluctuations has been reported for the alkali metals.

In this paper, we describe our first-principles calculations of the
contribution to the self energy arising from spin fluctuations. In
particular, we employ the spin-fluctuation self energy formalism
proposed by Kukkonen and Overhauser \cite{OverhauserKukkonen} which is
simpler than the T-matrix approach since it requires neither the
solution of a four-point equation nor a double counting correction. We
apply the theory to sodium and lithium, and find that the contribution
of spin fluctuations to the reduction of the occupied bandwidth is
small and, by itself, cannot explain experimental findings.  We also
carry out standard GW calculations and GW calculations with a
vertex-corrected charge susceptibility, with the latter resulting in a
larger reduction of the occupied bandwidth in agreement with
experiment and results in Refs. \cite{Northrup1,Northrup2}.

\emph{Methods}.---The properties of quasiparticles, such as their
energy or lifetime, can be measured in photoemission or tunneling
experiments. Mathematically, quasiparticle energies are given by the
positions of the poles of the single-particle Green's function and can
be determined by solving the Dyson equation
\begin{align}
 &\left( -\frac{\nabla^2}{2} + V_{\text{ion}}(\bm{r}) + V_H(\bm{r}) \right)
  \psi_n(\bm{r}) +  \nonumber \\ &\int
  d\bm{r}'\Sigma(\bm{r},\bm{r}',E_n)\psi_n(\bm{r}') = E_n \psi_n(\bm{r}),
  \label{eq:dyson}
\end{align}
where $V_{\text{ion}}$ and $V_H$ denote the external potential due the
ionic cores and the Hartree potential, respectively, $E_n$ and
$\psi_n$ are the quasiparticle energy and wave function, and $\Sigma$
denotes the electron self energy.

We approximate the self energy using
\cite{OverhauserZhu,VignaleSingwi}
\begin{align}
  \Sigma(\bm{r},\bm{r}',\omega) = i\int \frac{d\omega'}{2\pi}
  e^{-i\delta \omega'} G(\bm{r},\bm{r}',\omega-\omega')
  V_{\text{eff}}(\bm{r},\bm{r}',\omega'),
  \label{eq:sigma}
\end{align}
where $G$ and $V_{\text{eff}}$ denote the single-particle Green's
function and the effective interaction between electrons,
respectively, and $\delta=0^+$. We separate $V_{\text{eff}}$ and
subsequently $\Sigma$ into three contributions: a bare Coulomb term
$v(\bm{r},\bm{r}')=1/|\bm{r}-\bm{r}'|$ (resulting in the bare exchange
contribution to the self energy), a charge-fluctuation mediated
interaction $\delta W_C$ (resulting in a correlation contribution to
$\Sigma$ arising from charge fluctuations), and a spin-fluctuation
mediated interaction $\delta W_S$ (resulting in a spin-fluctuation
contribution to $\Sigma$).

In a standard GW calculation \cite{LouieHybertsen,HedinBook}, $\delta
W_S=0$ and $\delta W_C=v \chi^{\text{RPA}}_C v$ (suppressing all
arguments and integrals for clarity) are assumed, where
$\chi^{\text{RPA}}_C$ denotes the interacting charge susceptibility in
the RPA. We will refer to this approximation as GW$_{\text{RPA}}$.

Following Refs. \cite{Northrup1,Northrup2} we include vertex
corrections to the dielectric screening by calculating $\chi_C$ from
DFT, i.e. by solving $\chi_C = \chi_0 + \chi_0 (v+f_{xc}) \chi_C$,
where $\chi_0$ denotes the non-interacting Kohn-Sham susceptibility
and $f_{xc}(\bm{r},\bm{r}')=\delta^2 E_{xc}/(\delta \rho(\bm{r})\delta
\rho(\bm{r}'))$ with $E_{xc}$ and $\rho(\bm{r})$ being the
exchange-correlation energy within DFT and the electron density,
respectively. Note that $\chi_C$ is exact if we know the exact
$E_{xc}$. The simplest approximation to the exchange-correlation
energy is the local density approximation (LDA) \cite{KohnSham} and we
will refer to this approximation as GW$_{\text{LDA}}$. Inclusion of
vertex corrections in $\chi_C$ leads to a better satisfaction of the
Ward identity associated with particle conservation \cite{Hanke}.

The spin-fluctuation mediated interaction can be approximated
\cite{OverhauserZhu,OverhauserKukkonen} using
\begin{align}
  \delta W_S(\bm{r},\bm{r}',\omega) = 3\int d\bm{r}_1d\bm{r}_2 I_{xc}(\bm{r},\bm{r}_1) \chi_S(\bm{r}_1,\bm{r}_2,\omega) I_{xc}(\bm{r}_2,\bm{r}'),
\end{align}
where $\chi_S=\chi_0 + \chi_0 I_{xc} \chi_S$ denotes the interacting
spin susceptibility and $I_{xc}(\bm{r},\bm{r}') = \delta^2
E_{xc}/(\delta m(\bm{r})\delta m(\bm{r}'))$ with $m(\bm{r})$ denoting
the spin density. The factor of 3 results from the vector boson
nature of the spin fluctuations.

This intuitively appealing expression for the self energy arising from
spin fluctuations was first derived by considering the effective
interaction between electrons in a homogeneous electron gas including
exchange and correlation effects
\cite{OverhauserKukkonen}. Later\cite{VignaleSingwi} it was found that
the same expression may be obtained from an analysis of Feynman
diagrams \footnote{Refs. \cite{OverhauserKukkonen,VignaleSingwi}
  suggested further exchange and correlations effects may be included
  into the charge-fluctuation mediated interaction, $\delta W_C =
  (v+f_{xc}) \chi_C (v+f_{xc})$. In agreement with previous studies
  \cite{OverhauserZhu}, we find however that use of this interaction
  worsens the agreement with experiment compared to the
  GW$_{\text{LDA}}$ theory.}.

\begin{figure}
  \includegraphics[width=8.cm]{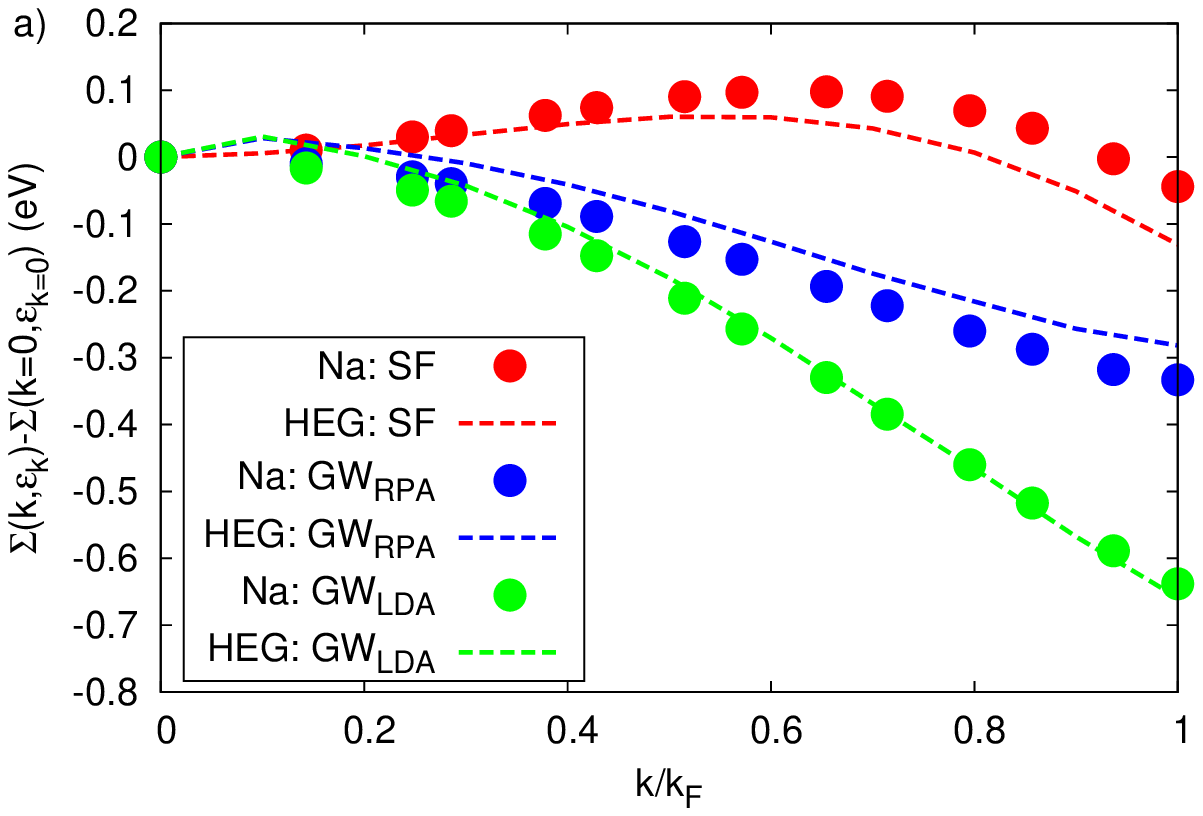} 
  \includegraphics[width=8.cm]{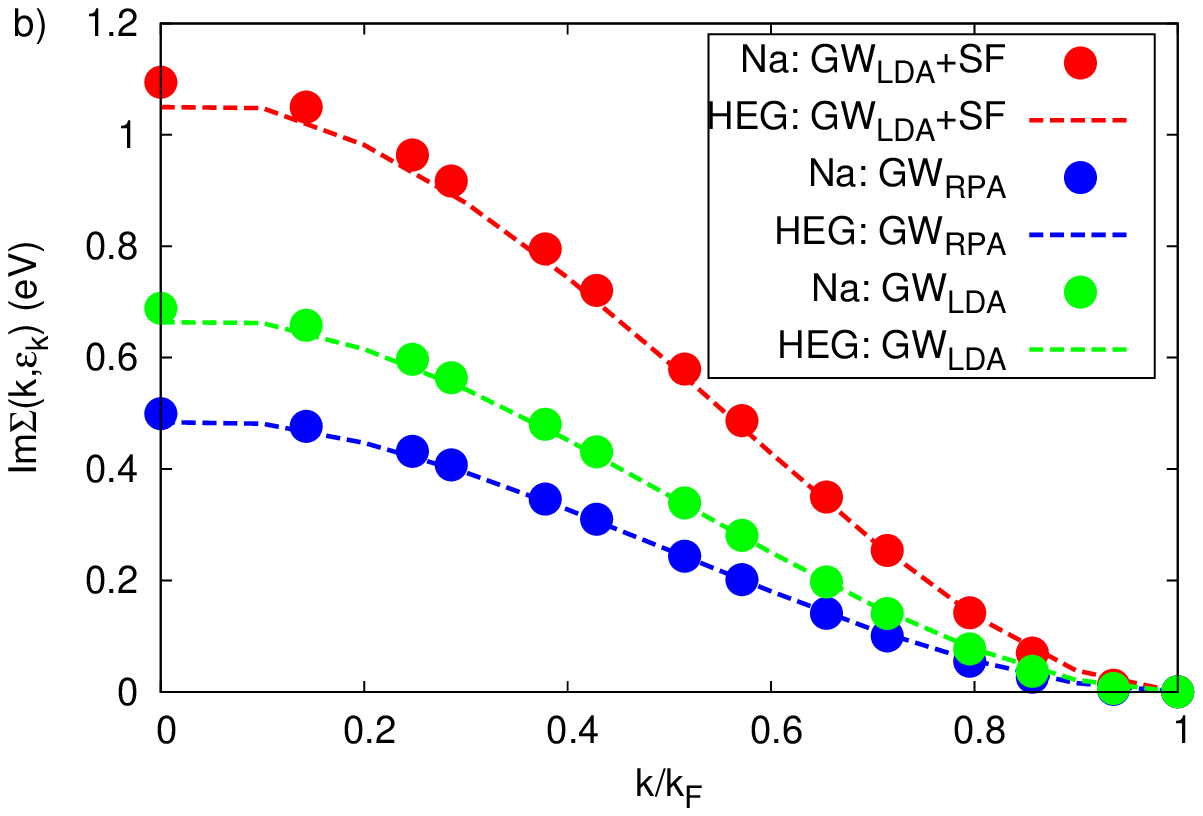}
  \caption{a) On-shell self-energy corrections for the occupied
    states in sodium. The dashed curves are results for the
    homogeneous electron gas (HEG) with $r_s=3.93$, while the circles
    are results from \emph{ab initio} calculations. The
    GW$_{\text{RPA}}$ theory (blue), GW$_{\text{LDA}}$ theory (green),
    and the spin-fluctuation self energy (red) are shown. b) Imaginary part of
    the on-shell self energy for the occupied states in sodium. The
    labels are the same as in a).}
  \label{fig:sodium}
\end{figure}

\emph{Computational details}.---We carry out DFT-LDA calculations
using a plane-wave basis and normconserving pseudopotentials as
implemented in the Quantum Espresso program package
\cite{QuantumEspresso}. Our plane wave cutoff is 30 Ry. For sodium we
choose the unit cell corresponding to $r_s=3.93$ and for lithium
corresponding to $r_s=3.26$ [$r_s$ is related to the valence charge
density $n$ via $n = 3/(4\pi (r_s a_B)^3_s)$ with $a_B$ being the Bohr
radius].  The self energy is calculated using the BerkeleyGW
\cite{BGWpaper} program package. For the calculation of the
susceptibilities and the self energies we use $16\times 16\times 16$
k-point sampling of the Brillouin zone. In our first-principles
calculations, we do not employ a generalized plasmon-pole model for
the interacting charge and spin susceptibilities, but sample these
quantities along the real frequency axis. We use fine sampling with a
step size of 0.1 eV up to a lower cutoff of 30 eV and then coarser
sampling up to 60 eV. A broadening of 0.15 eV is used as well as 30
empty states in the calculation of the dielectric matrix and the self
energy. In this work, we employ a one-shot procedure to calculate the
self energy. The effect of self-consistency on the occupied band width
of simple metals was investigated in Ref. \cite{Northrup1} and found to be
quite small. 

\emph{Sodium}.---The occupied bandwidth of sodium in DFT is 3.19
eV. Figure \ref{fig:sodium}(a) shows the self-energy correction
(evaluated ``on-shell'', i.e. at the mean-field energy) to the DFT-LDA
band structure from charge fluctuations (including the bare exchange)
and from spin fluctuations. We find a reduction of the occupied
bandwidth with charge fluctuations giving a significantly larger
contribution than spin fluctuations. In agreement with previous
calculations\cite{Northrup1,Northrup2}, we find that vertex
corrections in the charge susceptibility are very important
(increasing the bandwidth reduction by a factor of two compared to the
standard GW$_{\text{RPA}}$ result): standard GW$_{\text{RPA}}$ theory
gives a reduction of 0.31 eV, while GW$_{\text{LDA}}$ yields a
reduction of 0.63 eV resulting in an occupied bandwidth in good
agreement with the experimental findings \cite{Plummer1,Plummer2}. See
Table~\ref{table:sodium}.  The contribution from spin fluctuations to
the bandwidth reduction is very small, less than 0.1 eV.

To understand these results we observe that retaining only the bare
exchange contribution to the self energy results in a drastic
\emph{increase} of the occupied bandwidth by $3.30$~eV as expected
from usual Hartree-Fock theory. Inclusion of screening by charge
fluctuations has the opposite effect yielding a net reduction of the
bandwidth. The RPA underestimates the screening which explains the
larger bandwidth reduction in GW$_{\text{LDA}}$ theory
\cite{Northrup2}.

Figure \ref{fig:sodium}(a) also shows the result from a self-energy
calculation for the homogeneous electron gas (jellium) with $r_s=3.93$
corresponding to the valence charge density of sodium, $n=3/(4\pi [r_s
a_B]^3)$ with $a_B$ being the Bohr radius. The occupied bandwidth in
Hartree theory is 3.15 eV. This agrees very well with the \emph{ab
  initio} DFT-LDA result indicating that corrections caused by the
inhomogeneity of the crystalline potential are very
small. Fig.~\ref{fig:sodium} shows good agreement between the on-shell
self energies from the \emph{ab initio} calculation and jellium.

If the exact self energy was known, quasiparticle properties should be
calculated ``off-shell'', i.e. the quasiparticle equation,
Eq.~\eqref{eq:dyson}, should be solved and the self energy should be
evaluated at the quasiparticle energy. It has been argued
\cite{Rice,DasSarmaDyson}, however, that for approximate self energies
quasiparticle properties should be determined by evaluating the self
energy ``on the shell'' to avoid mixing different orders of
perturbation theory. Table~\ref{table:sodium} shows that off-shell
calculations result in a larger occupied bandwidth than on-shell
calculations. The ratio of the bandwidth reductions in off-shell and
on-shell calculations is approximately equal to the renormalization
constant $Z_{\bm{k}}=[1-\partial \Sigma_{\bm{k}}(E^{qp}_{\bm{k}})
/\partial \omega]^{-1}$, which is about 0.6 for the occupied states of
sodium.

According to Eq.~\eqref{eq:sigma}, the self energy should be
calculated using the interacting Green's function. To approximately
account for this self-consistency requirement we have shifted all
mean-field energies such that the resulting quasiparticle energy
agrees with the shifted mean-field energy \cite{HedinReview} at the
Fermi level. Table~\ref{table:sodium} shows that self-consistency only
leads to very small changes in the occupied bandwidth.

\begin{table}
  \setlength{\doublerulesep}{1\doublerulesep}
  \setlength{\tabcolsep}{2\tabcolsep}
    \caption{Occupied bandwidth of sodium obtained from on-shell and
      off-shell evaluations of the self energy. We also give results
      with approximate self-consistency (sc) achieved by shifting the
      mean-field energies. All energies are given in eV.}
  \begin{ruledtabular}
    
    \begin{tabular}{c | c c c }
       & on-shell & off-shell & off-shell + sc \\
      \hline
      GW$_{\text{RPA}}$        &  2.86 &  3.00  & 2.98 \\
      GW$_{\text{LDA}}$         & 2.55 &  2.83  & 2.80 \\
      GW$_{\text{LDA}}$+SF   & 2.51 &  2.85  & 2.78 \\
      \hline
      exp. \cite{Plummer1}                          & 2.5 &  & \\
      exp.  \cite{Plummer2}                         & 2.65 &&\\
      DFT-LDA                  & 3.19  & & \\
    \end{tabular}

  \end{ruledtabular}
  \label{table:sodium}
\end{table}

\begin{figure}
  \includegraphics[width=8.cm]{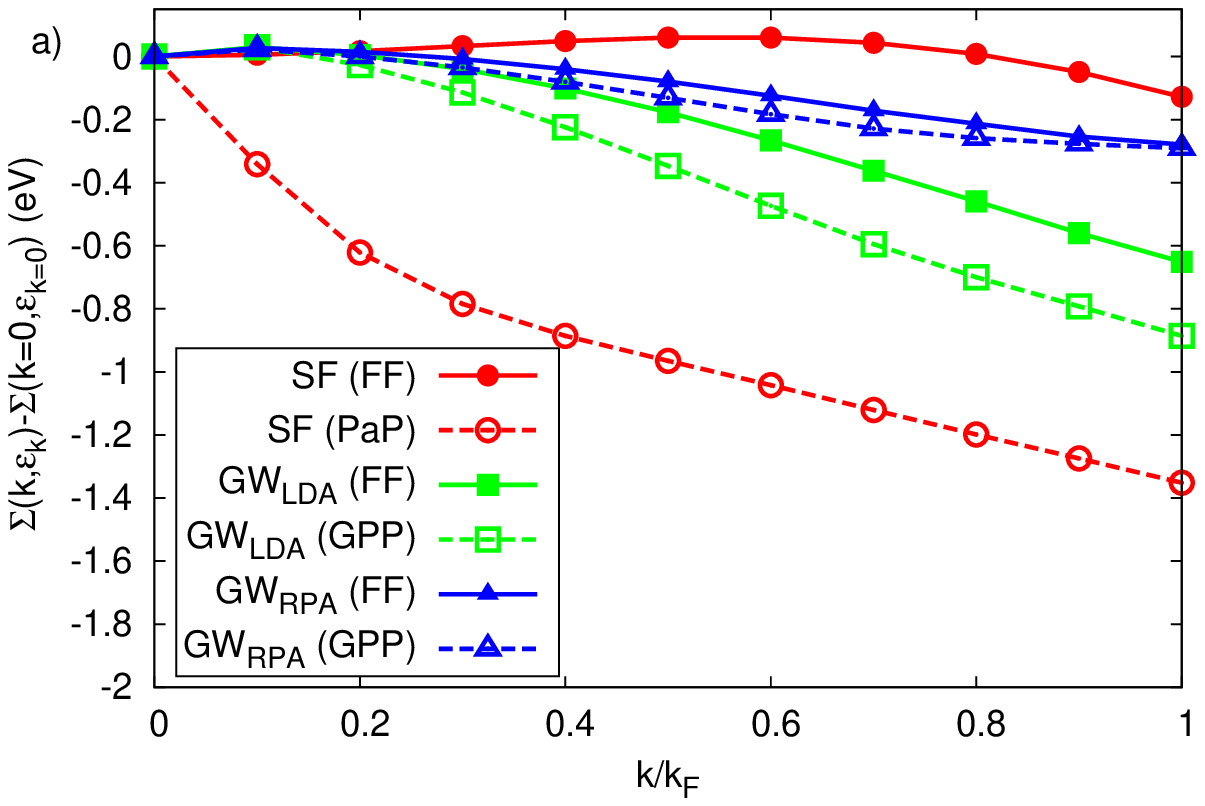}
  \includegraphics[width=8.cm]{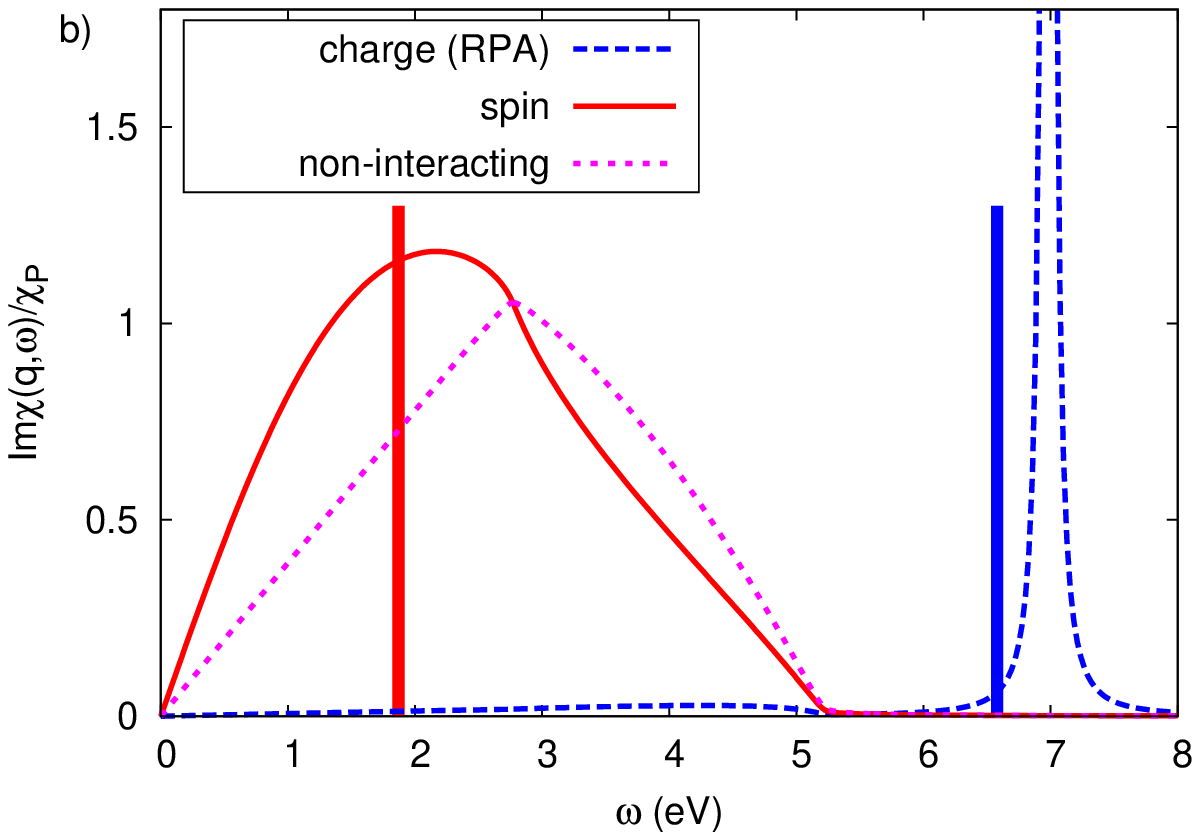}  
  \caption{a) Comparison of the on-shell self-energy corrections for
    the occupied states of jellium with $r_s=3.93$ (corresponding to
    sodium) from full-frequency and simplified calculations using the
    generalized plasmon-pole (GPP) and paramagnon-pole (PaP) models
    for the GW$_{\text{RPA}}$ self energy (blue), GW$_{\text{LDA}}$
    self energy (green), and spin-fluctuation self energy (red). b)
    Frequency-dependent imaginary parts of the charge (blue dashed
    curve), spin (red solid curve), and non-interacting (magenta
    dotted curve) susceptibilities for jellium at $q/q_F=0.6$. The
    susceptibilities are divided by the Pauli susceptibility
    $\chi_P=-k_F/\pi^2$. The vertical lines denote the locations of
    the delta-function peaks in a plasmon-pole model (blue)
    \cite{Lundqvist} and a paramagnon-pole model (red)
    \cite{OverhauserZhu}.}
  \label{fig:pp}
\end{figure}

We employed the jellium model to investigate the effect of additional
approximations to the self energy. In standard first-principles
GW$_{\text{RPA}}$ calculations, one sometimes employs a generalized
plasmon-pole model \cite{LouieHybertsen} to extend the static inverse
dielectric matrix to finite frequencies. In this model the imaginary
part of the inverse dielectric function for each $\bm{G}$ and
$\bm{G}'$ component is assumed to be a simple delta-function,
i.e. $\text{Im}\epsilon_{\bm{GG}'}^{-1}(\bm{q},\omega) \propto
\delta(\omega-\omega_{\bm{GG}'}(\bm{q}))$ with
$\omega_{\bm{GG}'}(\bm{q})$ denoting the effective plasmon
frequency. Fig.~\ref{fig:pp}(a) shows that the plasmon-pole model
reproduces the self-energy shifts arising from charge fluctuations
quite well. Zhu and Overhauser \cite{OverhauserZhu} employed a similar
\emph{paramagnon-pole model} to simplify the calculation of the
spin-fluctuation self energy: they assumed that the imaginary part of
the interacting spin susceptibility can be represented by a single
mode, the paramagnon. However, in contrast to the plasmon which cannot
decay into particle-hole pairs for small wave vectors, the paramagnon
has a linear acoustic-like dispersion (as determined by the f-sum rule
\cite{OverhauserZhu}) and can decay into particle-hole pairs:
Fig.~\ref{fig:pp}(b) shows the imaginary part of the spin, charge, and
non-interacting susceptibilities at $q/q_F = 0.6$. The charge
susceptibility has a sharp plasmon peak at $\sim 7$~eV that lies
outside the particle-hole continuum given by the non-interacting
susceptibility. The spin susceptibility has significant overlap with
the non-interacting susceptibility and exhibits a broad structure.  In
contrast to the charge susceptibility, the spin susceptibility is not
well represented by a single sharp mode.

As a consequence, the spin-fluctuation self energy with a
paramagnon-pole model gives very different results from the theory
without this approximation, see Fig.~\ref{fig:pp}(a). It results in a
drastic narrowing of the occupied bandwidth by $\sim 1.4$~eV. While
Zhu and Overhauser \cite{OverhauserZhu} empirically correct for the
finite lifetimes of the paramagnon, it is likely that their predicted
bandwidth narrowing of 0.7~eV is also spuriously large and caused by
the paramagnon-pole approximation.

Figure~\ref{fig:pp} also shows that for the GW$_{\text{LDA}}$ theory
the plasmon-pole approximation leads to a further reduction of the
occupied bandwidth by $\sim 0.2$~eV. The resulting bandwidth
(calculated ``off-shell'') agrees well with previous first-principles
results \cite{Northrup1,Northrup2}.

Figure~\ref{fig:sodium}(b) on the other hand shows that spin
fluctuations contribute significantly to the line width
[$2\text{Im}\Sigma_{\bm{k}}(\epsilon_{\bm{k}})$] of quasiparticles. In
particular, at the bottom of the band, $k=0$, the quasiparticle line
width is a factor of 2 larger than the GW$_{\text{RPA}}$ value when
vertex corrections in $\chi_C$ and spin-fluctuations are
included. This agrees well with the experimental findings
\cite{PlummerReview,Penn,Echenique2}.

\emph{Lithium.---} In contrast to sodium, the occupied bandwidth
resulting from Hartree theory applied to the homogeneous electron gas,
4.65 eV, is much larger than the value obtained in a DFT-LDA
calculation including the crystalline potential, 3.45 eV. This shows
that --- even though the Fermi surface is spherical to a high degree
\cite{Callaway} --- crystal effects are very important in lithium.

\begin{table}
  \setlength{\doublerulesep}{1\doublerulesep}
  \setlength{\tabcolsep}{2\tabcolsep}
    \caption{Occupied bandwidth of lithium obtained from on-shell and
      off-shell evaluations of the self energy. We also give results
      with approximate self-consistency (sc) achieved by shifting the
      mean-field energies. All energies are given in eV.}
  \begin{ruledtabular}
    
    \begin{tabular}{c | c c c }
       & on-shell & off-shell & off-shell + sc \\
      \hline
      GW$_{\text{RPA}}$        &  3.19 &  3.29  & 3.30 \\
      GW$_{\text{LDA}}$         & 2.83 &  3.09  & 3.10 \\
      GW$_{\text{LDA}}$+SF   & 2.87 &  3.24  & 3.08 \\
      \hline
      exp.  \cite{LiExp}                         & 2.86 &  & \\
      DFT-LDA                  &  3.45&  &  \\
    \end{tabular}

  \end{ruledtabular}
  \label{table:lithium}
\end{table}

Table~\ref{table:lithium} shows our results for the occupied bandwidth
of lithium. Again, the occupied bandwidth reduction in
GW$_{\text{RPA}}$ theory is too small to explain the experimental
finding. Adding vertex correction in $\chi_C$ yields good agreement
with experiment \cite{LiExp}, while spin fluctuations only lead to a
small change in the occupied bandwidth.

\emph{Conclusions}.---We have calculated the effect of spin and charge
fluctuations on quasiparticle excitations in alkali metals from first
principles. In contrast to previous calculations \cite{OverhauserZhu},
we find that spin fluctuations contribute little to the observed
bandwidth reduction compared to mean-field results. Instead, as
observed in Refs. \cite{Northrup1,Northrup2} inclusion of vertex
corrections in the dielectric screening gives agreement with
experimental bandwidths. Previous studies which included vertex
corrections in both the dielectric screening and the self energy found
that agreement with experiment worsens
\cite{MahanSernelius,Godby}. Other studies reported cancellations
between self-energy vertex corrections and selfconsistency effects
\cite{HolmBarth,HolmBarth2}. Further work is necessary to clarify this
issue.  We also find that spin fluctuations give an important
contribution to the line width and lifetime of the quasiparticle
excitations. We note that the first-principles framework presented
here can be applied to materials with d-electrons where spin
fluctuations are expected to play an essential role, such as pnictide
superconductors and ferromagnetic metals.

S. G. L. acknowledges support by a Simons Foundation Fellowship in
Theoretical Physics. This work was supported by NSF Grant
No. DMR10-1006184 (numerical simulations of the alkali metals) and by
the Director, Office of Science, Office of Basic Energy Sciences,
Division of Materials Sciences and Engineering Division, US Department
of Energy under Contract No. DE-AC02-05CH11231 (software development
of electron correlation effects).

\bibliography{paper}
\end{document}